\documentclass[conference,10pt]{IEEEtran}
\usepackage{cite}
\usepackage{epsfig}
\usepackage{epstopdf}
\usepackage{graphicx}
\usepackage[dvipsnames]{xcolor}
\usepackage{tikz}
\usetikzlibrary{spy,backgrounds,arrows,positioning,shapes.geometric,circuits.logic.US}
\usepackage{pgfplots}
\usepackage{multirow}
\usepackage{upgreek}
\usepackage{amssymb}
\usepackage{amsmath}
\usepackage{cases}
\usepackage{url}
\usepackage{pifont}
\usepackage{bm}
\usepackage{amsthm}

\usepackage{tabularx}
\usepackage{booktabs}
\usepackage{filecontents}
\usepackage{subcaption}
\newcolumntype{Y}{>{\centering\arraybackslash}X}

\theoremstyle{definition} 
\theoremstyle{definition} 
\theoremstyle{definition} 
\theoremstyle{definition}

\usepackage[linesnumbered, ruled, vlined, noend]{algorithm2e}

\usepackage{algorithmic}

\usepackage{array}

\newcommand{\fixme}[2]{\ifx&#2&{\leavevmode\color{red}#1}\else{\leavevmode\color{red}FIXME\{}#1{\leavevmode\color{red}\}}\footnote{{\leavevmode\color{red}#2}}\PackageWarning{Fixme}{#1: #2}\fi}

\newcommand{\newstuff}[2]{\ifx&#2&{\leavevmode\color{blue}#1}\else{\leavevmode\color{blue}FIXME\{}#1{\leavevmode\color{blue}\}}\footnote{{\leavevmode\color{blue}#2}}\PackageWarning{Newstuff}{#1: #2}\fi}

\title{Input-distribution-aware parallel decoding\\ of block codes}

\author{\IEEEauthorblockN{Carlo~Condo, Alex~Nicolescu}
\IEEEauthorblockA{Infinera Corporation}}%\\
\begin{document}

\maketitle
\begin{abstract}

Many channel decoders rely on parallel decoding attempts to achieve good performance with acceptable latency. 
However, most of the time fewer attempts than the foreseen maximum are sufficient for successful decoding.  
Input-distribution-aware (IDA) decoding allows to determine the parallelism of polar code list decoders by observing the distribution of channel information.
In this work, IDA decoding is proven to be effective with different codes and decoding algorithms as well. 
Two techniques, M-IDA and MD-IDA, are proposed: they exploit the sampling of the input distribution inherent to particular decoding algorithms to perform low-cost IDA decoding. 
Simulation results on the decoding of BCH codes via the Chase and ORBGRAND algorithms show that they perform at least as well as the original IDA decoding, allowing to reduce run-time complexity down to $17\%$ and $67\%$ with minimal error correction degradation.

\end{abstract}

%\begin{IEEEkeywords}
%
%\end{IEEEkeywords}

\IEEEpeerreviewmaketitle

\section{Introduction} \label{sec:intro}

Forward error correction (FEC) is a critical component of modern communication and storage systems. 
Efficient implementation of FEC decoders is of paramount importance in modern applications, given the high computational cost of decoding powerful channel codes and the growing emphasis on low-power technology.

A large number of high performance FEC schemes and decoding algorithms rely on a trial and evaluation loop, in which the decoding is attempted under some conditions, and in case of failure, the decoding conditions are changed to give the decoder better chances of succeeding. 
For example, fountain codes can generate extra parity symbols  \cite{Fountain}, polar code flip decoders sequentially change the decoding decision on low-reliability bits \cite{SCFLIP}, and low-density parity-check code decoders can inject noise and perform additional iterations to escape from decoding failure \cite{NANLDPC}. 
These techniques, while powerful, incur variable and possibly very long latency, something that is not always sustainable in practical systems.

To avoid such drawback, many decoding algorithms for block codes are based on parallel decoding paths, each taking different decisions or working on a modified version of the information received from the channel \cite{TalSCL,SCANL,Chase,ORBGRAND}. 
The higher the number of paths, the more the decoding process is able to cater to complex, less probable noise patterns. 
This means that most noise occurrences can be corrected with very few paths. 
It would be beneficial for practical decoders to be able to identify such cases, and to deactivate some of the parallel paths to reduce power consumption.

In \cite{IDASCL}, input-distribution-aware (IDA) successive cancellation list (SCL) \cite{TalSCL} decoding of polar codes has been proposed, a technique that allows to decide which list size to adopt by observing the input of the decoder.  
This technique guarantees fixed, short decoding latency and allows hardware SCL decoders to dynamically decrease the list size before each decoding by shutting down part of the internal parallelism. 

In this work, the generality of IDA decoding is proven by evaluating its effectiveness on different types of codes and decoding algorithms. 
Two low-complexity incarnations of IDA decoding are proposed.  
These rely on inherent characteristics of decoding algorithms to decrease the implementation complexity of IDA without impacting its performance.
 
\section{Preliminaries} \label{sec:prel}

\subsection{Definitions}\label{subsec:defs}

In this work, it is assumed that the decoder receives soft values from the channel in the form of logarithmic likelihood ratios (LLRs). 
Let us call the received LLR vector $\mathbf{y}$, and $\mathbf{\hat{x}}$ the vector of hard decisions taken on $\mathbf{y}$:
\begin{equation}\label{eqn:HD}
\hat{x}_{i}=\left\{
  \begin{array}{@{}ll@{}}
    0, & \text{when } y_i \geq 0 ; \\
    1, & \text{otherwise.}
  \end{array}\right.
\end{equation}
Moreover, let us define as $\mathbf{\tilde{y}}$ the received vector $\mathbf{y}$ sorted in ascending order of absolute value.
Analysis and simulations all consider additive white Gaussian noise (AWNG) channel with binary phase shift keying (BPSK) modulation, for which LLRs can be computed as $\boldsymbol{y}=2/\sigma^2 \cdot (1-2\boldsymbol{x} + E )$, where $\sigma$ is the standard deviation of the noise distribution, and the random variable $E\sim \mathcal{N}(0,\sigma^2)$ has equiprobable sign and magnitude that increases with decreasing probability. 

Finally, throughout the paper, ``\textbf{correct decoding}" and ``\textbf{successful decoding}" refer to the event where a decoder corrects the errors accrued during transmission, and is able to successfully retrieve the information bits that were encoded at the transmitter side.

\subsection{Previous work}\label{subsec}

\begin{figure}[t!]
  \centering
  \begin{tikzpicture}
  \pgfplotsset{
    label style = {font=\fontsize{8pt}{7.2}\selectfont},
    tick label style = {font=\fontsize{6pt}{7.2}\selectfont}
  }

\begin{axis}[
	scale = 1,
    xlabel={LLR value}, xlabel style={yshift=-0.2em},
    ylabel={Average number of LLRs}, ylabel style={yshift=-0.75em},
    grid=both,
    ymajorgrids=true,
    xmajorgrids=true,
    grid style=dashed,
    mark options=solid,
    ybar interval,
    ymax = 256,
    ymin = 0,
    width=1\columnwidth, height=5.8cm,
    thick,
	xmin=1,
	xmax=12,
	xtick={1,2,3,4,5,6,7,8,9,10,11,12},
    xticklabels={$<$0.5,$<$1.0,$<$1.5,$<$2.0,$<$2.5,$<$3.0,$<$3.5,$<$4.0,$<$4.5,$<$5.0,$<$5.5,$<$6.0},
    legend style={
      anchor={center},
      cells={anchor=west},
      mark options=solid,
      column sep= 2mm,
      font=\fontsize{7pt}{7.2}\selectfont,
    },
    legend to name=LLRdist,
    legend columns=3,
]

\addplot[
	color=CornflowerBlue,
	fill,
]
table {
1 16.13	
2 32.708	
3 50.033	
4 68.297	
5 87.462	
6 107.3	
7 127.37
8 147.11	
9 165.96
10 183.39
11 198.97
12 212.43
};
\addlegendentry{$L$=1}

\addplot[
    color=blue,
	fill,
]
table {
1   17.075	
2   34.57	
3   52.608	
4   71.315	
5   90.612	
6   110.39	
7   130.3	
8   149.75	
9   168.26	
10  185.32	
11  200.5	
12  213.59	
};
\addlegendentry{$L$=2}

\addplot[
    color=Black,
	fill,
]
table {
1  17.55
2  35.407
3  53.812
4  72.894
5  92.525
6  112.33
7  131.98
8  151.22
9  169.56
10 186.38
11 201.45
12 214.35
};
\addlegendentry{$L$=4}

\addplot[
    color=Gray,
	fill,
]
table {
1  17.92	
2  36.051	
3  54.747	
4  74.073	
5  93.607	
6  113.71	
7  133.52	
8  152.81	
9  170.94	
10 187.37	
11 202.29	
12 214.98	
};
\addlegendentry{$L$=8}

\addplot[
    color=Bittersweet,
	fill,
]
table {
1   18.297
2   36.933
3   55.527
4   74.908
5   95.1	
6   114.6
7   134.68
8   153.84
9   171.93
10  188.65
11  203.41
12  215.95
};
\addlegendentry{$L$=16}

\addplot[
color=BurntOrange,
	fill,
]
table {
1  18.769
2  37.824
3  56.324
4  75.769
5  96.056
6  116.3
7  136.13
8  154.83
9  172.57
10 188.71
11 203.12
12 216.07
};
\addlegendentry{$L$=32}

\end{axis}
\end{tikzpicture}
  \\
 \ref{LLRdist}
  \caption{LLR distribution for a polar code with $N$=256 at $E_b/N_0$=2.4 dB with $L$ necessary decoding paths.}
  \label{fig:LLRdist}
\end{figure}
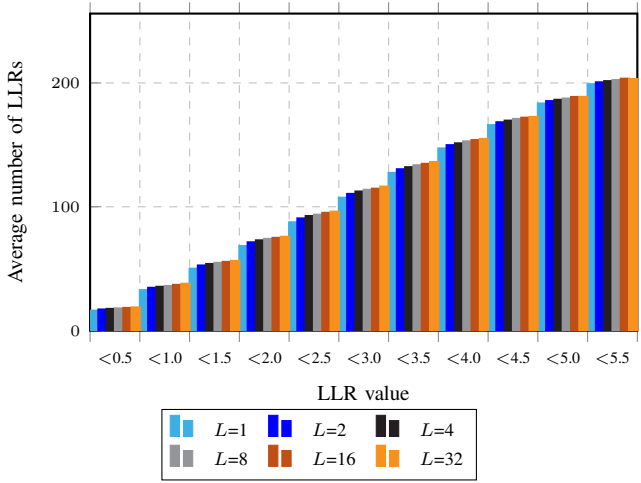

In \cite{IDASCL}, input-distribution-aware (IDA) successive cancellation list (SCL) decoding of polar codes was proposed. 
It relies on the observation of the link between the distribution of soft values received from the channel, usually LLRs, and the number of concurrent decoding paths employed by SCL to achieve correct decoding. 
Figure \ref{fig:LLRdist}, taken from \cite{IDASCL}, shows that the average distribution of channel LLRs $y$ for a polar code of length $N=256$ changes with the number of parallel decoding attempts $L$ necessary to correctly decode it. 
It was shown that the parallelism $L_{high}$ of a decoder can be reduced on-the-fly to $L_{low}<L_{high}$, with a limited cost in block error rate (BLER) and substantial complexity reduction.
The channel LLRs that are lower or equal than a threshold $\gamma$ are counted; if their number is lower than a threshold $\phi$, then the decoding is attempted with $L_{low}$, otherwise $L_{high}$ is used. 
This operation can be implemented at a low cost in practical decoders. 
The effectiveness of IDA decoding was proven over a variety of code lengths, rates and channel conditions. 
The run-time complexity of such a decoder, intended as the percentage of active decoding paths with respect to the total, is computed as
\begin{equation} \label{eq:comp_SL}
100 \cdot \frac{\delta\cdot L_{low}+(1-\delta)\cdot L_{high}}{L_{high}}~,	
\end{equation}
where $\delta$ is the fraction of times that $L_{low}$ is chosen over $L_{high}$.

\section{IDA decoding}\label{sec:IDA}

IDA decoding has been applied to SCL decoding of polar codes so far; however, it is a flexible technique, and it is not exclusive to polar codes or SCL decoding. 
Any decoder using parallel decoding paths at the end of which a decoding candidate is chosen, can benefit from the IDA approach.
As an example, let us consider Bose-Chaudhuri-Hocquenghem (BCH) codes, that are polynomial block codes widely used in communications, storage and cryptography applications. 
A BCH code of length $N$, with $K$ information bits, capable of correcting $t$ errors, is identified as BCH($N$,$K$,$t$). 
One of the most successful and widespread decoding algorithms for such codes is the Chase decoder \cite{Chase}. 
Chase decoding takes the hard decision of the input LLRs, but it also identifies the $P$ bits associated to the LLRs with the smallest magnitude, i.e. $\tilde{y}_i$ with $0\le i< P$. 
It then creates the set of $2^P$ test patterns where all combinations of these bits are flipped. 
The resulting $2^P$ vectors are then decoded, in parallel, through bounded distance decoding. 
However, for a decoder that considers $P_{high}$ LLRs, the vast majority of the  decoding processes can be successfully carried out with $P_{low}<P_{high}$.
This can be seen in Table \ref{tab:minP}, that reports the percentage of cases for which a specific $P_{low}$ is sufficient for successful decoding, given $P_{high}$=6 at $E_b/N_0$=6.5dB and a BCH(255,239,2) with generator polynomial $g_C(x)$=0x18DED defined on GF($2^8$) with a field generator polynomial $g_F(x)$=0x171. 
Chase decoding does not change the guaranteed $t$-error correction capability of BCH codes, but can potentially correct up to $P$ additional errors, if they are associated with the least reliable LLRs. 
The distribution Table \ref{tab:minP} is thus dependent on the number of errors in the codeword and on their average LLR. 
At lower $E_b/N_0$, higher $P_{low}$ will be necessary with increased probability.

Similar uneven decoding requirement distributions can be observed with other code types and decoding algorithms; without loss of generality, the remainder of this work uses BCH codes and considers the aforementioned Chase decoder and the recently proposed ordered reliability bits guessing random additive noise decoding (ORBGRAND).

\begin{table}[t!]
  \caption{Percentage of cases for $P_{low}$ requirement in Chase decoding of BCH(255,239,2), $P_{high}=6$, $E_b/N_0$=6.5dB.}
	\centering
	\begin{tabular}{|c|c|}
		\hline
		$P_{low}$ & Requirement \%  \\
   		\hline
		0  	 & $96.54\%$		\\
		1	&	$2.55\%$		\\
		2	&	$0.66\%$			\\
		3	&	$0.18\%$			\\
		4	&	$0.05\%$			\\
		5	&	$0.01\%$				\\
		6	&	$<0.01\%$			\\
		\hline	
	\end{tabular}
\label{tab:minP}
\end{table}

\subsection{M-IDA and MD-IDA decoding}

Observing the distribution of input LLRs is a general approach that allows to decrease the parallelism of many decoders. 
Some decoding algorithms, however, inherently perform a sampling of the input distribution, and perform parallel decoding on the basis of a subset of the input.
This feature can be exploited to enable IDA decoding at a lower complexity cost than the observation of the full channel LLR distribution.
It is the case of Chase decoding, that has been addressed earlier in this Section, and ORBGRAND \cite{ORBGRAND}, a universal decoder that evolves from GRAND \cite{GRAND}. 
ORBGRAND attempts to guess the noise vector that was applied during transmission, by considering nPat bit-flipping patterns and applying them to $\mathbf{\hat{x}}$.
Bit flipping patterns are determined offline based on their logistic weight $w$, i.e the sum of the indices of flipped bits. 
For example, the pattern flipping the least reliable bit has $w=1$, the one flipping the second least reliable has weight $w=2$, the one flipping both bits has weight $w=1+2=3$, and so on. 
The nPat patterns with the smallest weight are used at decoding time, after $\mathbf{\tilde{y}}$ is available and the actual positions of the least reliable bits in the given $\mathbf{\hat{x}}$ is determined \cite{ORBGRAND}.

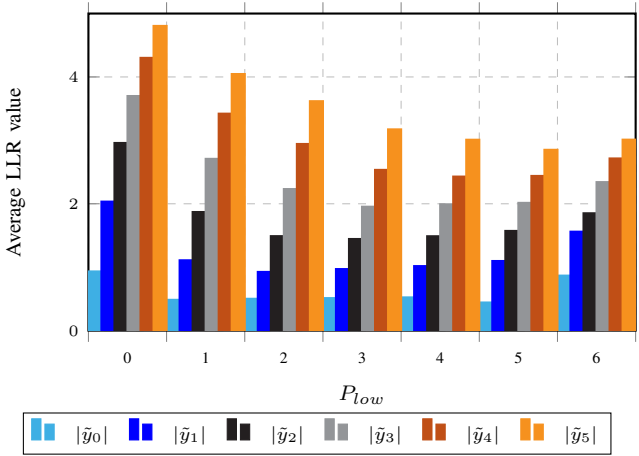
\begin{figure}[t!]
  \centering
  \begin{tikzpicture}
  \pgfplotsset{
    label style = {font=\fontsize{8pt}{7.2}\selectfont},
    tick label style = {font=\fontsize{6pt}{7.2}\selectfont}
  }

\begin{axis}[
	scale = 1,
    xlabel={$P_{low}$}, xlabel style={yshift=-0.2em},
    ylabel={Average LLR value}, ylabel style={yshift=-0.75em},
    grid=both,
    ymajorgrids=true,
    xmajorgrids=true,
    grid style=dashed,
    mark options=solid,
    ybar interval,
    ymax = 5,
    ymin = 0,
    width=1\columnwidth, height=5.8cm,
    thick,
	xmin=1,
	xmax=8,
	xtick={1,2,3,4,5,6,7,8},
    xticklabels={0,1,2,3,4,5,6,7},
    legend style={
      anchor={center},
      cells={anchor=west},
      mark options=solid,
      column sep= 2mm,
      font=\fontsize{7pt}{7.2}\selectfont,
    },
    legend to name=LRPdist,
    legend columns=6,
]

\addplot[
	color=CornflowerBlue,
	fill,
]
table {
1 0.932623495
2 0.484957188
3 0.501516826
4 0.512073798
5 0.523371526
6 0.442727333
7 0.866796667
8 0
};
\addlegendentry{$|\tilde{y}_0|$}

\addplot[
    color=blue,
	fill,
]
table {
1  2.033720643
2  1.108445678
3  0.924945725
4  0.970066245
5  1.016917457
6  1.097225556
7  1.559723333
8 0
};
\addlegendentry{$|\tilde{y}_1|$}

\addplot[
    color=Black,
	fill,
]
table {
1 2.958556715
2 1.869774295
3 1.489178976
4 1.444279939
5 1.487677143
6 1.571568889
7 1.849866667
8 0
};
\addlegendentry{$|\tilde{y}_2|$}

\addplot[
    color=Gray,
	fill,
]
table {
1 3.698766763 
2 2.708181521 
3 2.231699232 
4 1.953353374 
5 1.991522 
6 2.014058889
7 2.3413 
8 0
};
\addlegendentry{$|\tilde{y}_3|$}

\addplot[
    color=Bittersweet,
	fill,
]
table {
1 4.30069454  
2 3.420816066 
3 2.942688703 
4 2.534220307 
5 2.428531429 
6 2.438677778
7 2.7144 
8 0
};
\addlegendentry{$|\tilde{y}_4|$}

\addplot[
color=BurntOrange,
	fill,
]
table {
1 4.804275051
2 4.045739011
3 3.617222833
4 3.172987117
5 3.009648571
6 2.850733333
7 3.010833333
8 0
};
\addlegendentry{$|\tilde{y}_5|$}

\end{axis}
\end{tikzpicture}
  \\
 \ref{LRPdist}
  \caption{Distribution of average $|\tilde{y}_i|$ given $P_{low}$, BCH(255,239,2), $E_b/N_0=6.5$.}
  \label{fig:LRPdist}
\end{figure}

Both Chase and ORBGRAND decoders perform parallel decoding attempts based on a subset of LLRs with low reliability, whose cardinality is $P$ in case of Chase, and $w_{\rm nPat}$ in case of ORBGRAND. 
This subset constitutes a sensitive sample of the total channel LLR distribution, and a clear link can be observed between the first $P$ entries of $\mathbf{\tilde{y}}$ and the degree of parallelism required for correct decoding, much like in Fig. \ref{fig:LLRdist}. 
%In Chase decoding, let us define as $P_{low}$ the minimum $P$ required for successful decoding. 
Figure \ref{fig:LRPdist} plots the distribution of the average $|\mathbf{\tilde{y}}|$ as the required $P_{low}$ for correct decoding changes in case of BCH(255,239,2). 
Similar observations can be made for ORBGRAND in relation to the required $w_{{\rm nPat}_{\rm low}}$, and consequently nPat$_{low}$.
Based on this trend, it is possible to devise a simplified IDA decoding that relies on the observation of one of the entries of $\mathbf{\tilde{y}}$ only.
To decide which one, it can be seen that $|\tilde{y}_i|$ tends to decrease as $P_{low}$ increases, until $P_{low}=i$, whereas $|\tilde{y}_i|$ starts to increase. 
This is due to the fact that since most correctable error events are due to errors associated with low-reliability LLRs, a decoder requiring $P_{low}$ for successful decoding will be likely to observe a clustering of small LLRs $|\tilde{y}_i|$ with $i<P_{low}$ only. 
If $|\tilde{y}_i|$ with $i>=P_{low}$ were small as well, it would be less likely for the errors to be clustered in the first $P_{low}$ entries of $|\mathbf{\tilde{y}}|$.
Consequently, to decide between $P_{low}$ and $P_{high}$, the most stable choice is to observe $|\tilde{y}_{P_{high}-1}|$.
The same is true in case of $w_{\rm nPat}$ for ORBGRAND.
This new approach to IDA decoding is called M-IDA, being based on the magnitude of a single LLR. 
If $|\tilde{y}_{P_{high}-1}|$ ($|\tilde{y}_{w_{\rm nPat_{high}}-1}|$) is higher than a threshold $\gamma_M$, then the decoding is attempted with $P_{low}$ (nPat$_{low}$), otherwise $P_{high}$ (nPat$_{high}$) is used.

\begin{figure}[t!]
  \centering
  \begin{tikzpicture}
  \pgfplotsset{
    label style = {font=\fontsize{8pt}{7.2}\selectfont},
    tick label style = {font=\fontsize{6pt}{7.2}\selectfont}
  }

\begin{axis}[
	scale = 1,
    xlabel={$P_{low}$}, xlabel style={yshift=-0.2em},
    ylabel={Average value}, ylabel style={yshift=-0.75em},
    grid=both,
    ymajorgrids=true,
    xmajorgrids=true,
    grid style=dashed,
    mark options=solid,
    %ybar interval,
    ymax = 5,
    ymin = 0,
    width=1\columnwidth, height=5.8cm,
    thick,
	xmin=1,
	xmax=7,
	xtick={1,2,3,4,5,6,7,8},
    xticklabels={0,1,2,3,4,5,6,7},
    legend style={
      anchor={center},
      cells={anchor=west},
      mark options=solid,
      column sep= 2mm,
      font=\fontsize{7pt}{7.2}\selectfont,
    },
    legend to name=minmax,
    legend columns=3,
]

\addplot[
	color=CornflowerBlue,
	 mark=square,
    thick,
    mark size=3,
%	fill,
]
table {
1 3.871651556
2 3.560781823
3 3.115706007
4 2.660913319
5 2.486277045
6 2.408006
7 2.144036666
8 0
};
\addlegendentry{$|\tilde{y}_5|$ - $|\tilde{y}_0|$}

\addplot[
    color=blue,
	mark=o,
    thick,
    mark size=3,
%	fill,
]
table {
1  3.368071045
2  2.935858878
3  2.441171877
4  2.022146509
5  1.905159903
6  1.995950445
7  1.847603333
8 0
};
\addlegendentry{$|\tilde{y}_4|$ - $|\tilde{y}_0|$}

\addplot[
    color=Black,
    mark=x,
    thick,
    mark size=3,
%	fill,
]
table {
1 2.766143268
2 2.223224333
3 1.730182406
4 1.441279576
5 1.468150474
6 1.571331556
7 1.474503333
8 0
};
\addlegendentry{$|\tilde{y}_3|$ - $|\tilde{y}_0|$}

\addplot[
    color=Gray,
	    mark=triangle,
    thick,
    mark size=3,
%	fill,
]
table {
1 2.02593322
2 1.384817107
3 0.98766215
4 0.932206141
5 0.964305617
6 1.128841556
7 0.98307
8 0
};
\addlegendentry{$|\tilde{y}_2|$ - $|\tilde{y}_0|$}

\addplot[
    color=Bittersweet,
	    mark=+,
    thick,
    mark size=3,
%	fill,
]
table {
1 1.101097148
2 0.62348849
3 0.423428899
4 0.457992447 
5 0.493545931 
6 0.654498223
7 0.692926666
8 0
};
\addlegendentry{$|\tilde{y}_1|$ - $|\tilde{y}_0|$}

\end{axis}
\end{tikzpicture}
  \\
 \ref{minmax}
  \caption{Distribution of the difference between average $|\tilde{y}_i|$, $1 \le i \le 5$ and $|\tilde{y}_0|$, given $P_{low}$, BCH(255,239,2), $E_b/N_0=6.5$.}
  \label{fig:minmax}
\end{figure}
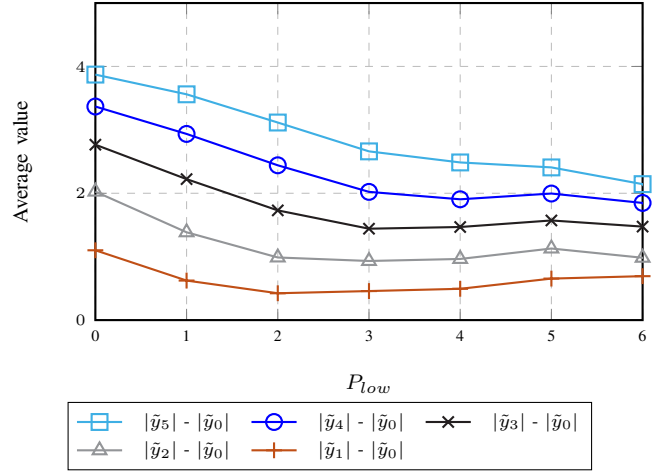

Following the same reasoning, an alternative metric to identify the necessary degree of parallelism is the difference between two elements of $|\mathbf{\tilde{y}}|$. 
It can give a measure of how clustered a set of LLRs are, and thus how likely it is that the decoder requires a high number of decoding attempts to succeed. 
Figure \ref{fig:minmax} plots the average difference between $|\tilde{y}_i|$, $1 \le i \le 5$ and $|\tilde{y}_0|$, for BCH(255,239,2) and Chase decoding, as the $P_{low}$ required for successful decoding changes. 
It can be observed that the the LLR difference is less affected by the trend inversion seen on the distribution of single LLRs in Figure \ref{fig:LLRdist},
in particular $|\tilde{y}_{P_{high}-1}|-|\tilde{y}_0|$, and can be used to implement a more conservative version of IDA decoding.
This is called MD-IDA, as it relies on the difference in magnitude of two LLRs. 
If $|\tilde{y}_{P_{high}-1}|-|\tilde{y}_0|$ ($|\tilde{y}_{w_{\rm nPat_{high}}-1}|-|\tilde{y}_0|$) is higher than a threshold $\gamma_{MD}$, then the decoding is attempted with $P_{low}$ (nPat$_{low}$), otherwise $P_{high}$ (nPat$_{high}$) is used. 

As Chase decoding requires all the $2^P$ attempts to be completed to be able to select the best candidate, both fully parallel and partially parallel implementations can benefit from IDA decoding \cite{HW_CHASE,HW_CHASE2}. 
On the other hand, test patterns in ORBGRAND are associated to a weight known at design time. 
Consequently, if they are applied in ascending order of weight, the algorithm can be stopped as soon as a valid codeword is obtained. 
However, ORBGRAND relies on a high number of simple operations; this approach would lead to extremely long worst case latency. 
While currently no implementations of ORBGRAND have been published, the only hardware GRAND decoder \cite{HW_GRAND} relies on a degree of parallelism of $N$, increasing the Hamming weight of the applied test patterns over subsequent time steps. 
With some modification to cater to logistic weights, the same concept can be applied to ORBGRAND, and such an architecture can exploit IDA decoding to reduce its power consumption.

%\fixme{Add max LRP and AVG LRP to tables}{}
%max and avg LRP decrease as the $LRP_{low}$ increases

\section{Results}\label{sec:results}

%\fixme{Introductory stuff, definition of legend}{}

In this Section are reported simulation results evaluating the effectiveness of the proposed techniques.
%The considered channel is AWGN with binary phase shift keying (BPSK) modulation, for which LLRs can be computed as $\boldsymbol{y}=2/\sigma^2 \cdot (1-2\boldsymbol{x} + E )$, where $\sigma$ is the standard deviation of the noise distribution, and the random variable $E\sim \mathcal{N}(0,\sigma^2)$ has equiprobable sign and magnitude that increases with decreasing probability. 
BLER and complexity (\ref{eq:comp_SL}) are computed on the BCH(255,239,2) code defined in Section \ref{sec:IDA} under the conditions presented in Section \ref{subsec:defs}.

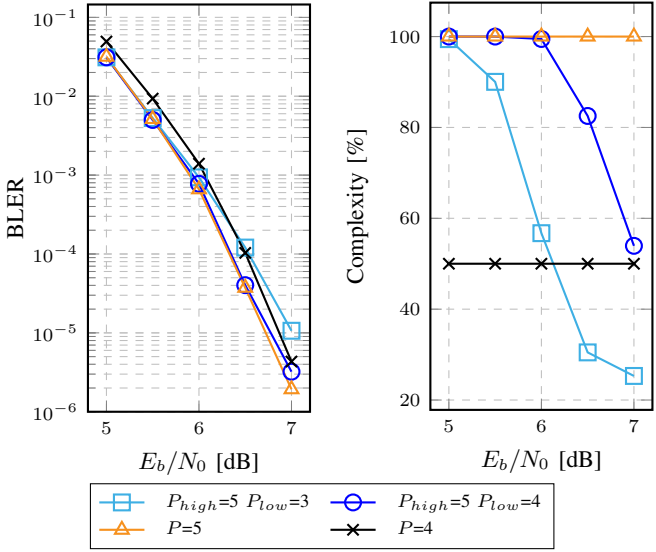
\begin{figure}[t!]
   \centering
    \begin{minipage}{.25\textwidth}
        \centering
%        		  \vspace{-12pt}
		  \begin{tikzpicture}
  \pgfplotsset{
    label style = {font=\fontsize{9pt}{7.2}\selectfont},
    tick label style = {font=\fontsize{7pt}{7.2}\selectfont}
  }

\begin{axis}[
	scale = 1,
    ymode=log,
    xlabel={$E_b/N_0$ [\text{dB}]}, xlabel style={yshift=0.4em},
    ylabel={BLER}, ylabel style={yshift=-0.75em},
    grid=both,
    ymajorgrids=true,
    xmajorgrids=true,
    grid style=dashed,
    mark options=solid,
    width=1\columnwidth, height=7cm,
    thick,
        ymin=1e-6,
    mark size=3,
    legend style={
      anchor={center},
      cells={anchor=west},
      mark options=solid,
      column sep= 2mm,
      font=\fontsize{7pt}{7.2}\selectfont,
    },
    legend to name=BLER_OPTIMAL_LLR,
    legend columns=2,
]

\addplot[
    color=CornflowerBlue,
    mark=square,
    thick,
    mark size=3,
]
table {
5       0.0310945	
5.5       0.0052815	
6      0.00093629	
6.5      0.00012063	
7     1.0639e-005	
};
\addlegendentry{$P_{high}$=5 $P_{low}$=3}
%4.5, 7

\addplot[
    color=blue,
    mark=o,
    thick,
    mark size=3,
]
table {
5       0.0310945	
5.5       0.0050015	
6      0.00077572	
6.5     4.0254e-005	
7       3.22e-006	
};
\addlegendentry{$P_{high}$=5 $P_{low}$=4}
%7.5, 12

\addplot[
    color=BurntOrange,
	mark=triangle,
    thick,
    mark size=3,
]
table {
5.0     0.0316556	
5.5       0.0051414	
6.0    0.00066028	
6.5     3.7284e-005	
7.0   1.9278e-006	
};
\addlegendentry{$P$=5}

\addplot[
    color=black,
    mark=x,
    thick,
    mark size=3,
]
table {
5.0     0.0493583	
5.5       0.0093266	
6.0     0.0013959	
6.5      0.00010328	
7.0   4.3302e-006		 
};
\addlegendentry{$P$=4}

\end{axis}
\end{tikzpicture}
		        \end{minipage}%
    \begin{minipage}{0.25\textwidth}
        \centering
		  \begin{tikzpicture}
  \pgfplotsset{
    label style = {font=\fontsize{9pt}{7.2}\selectfont},
    tick label style = {font=\fontsize{7pt}{7.2}\selectfont}
  }

\begin{axis}[
	scale = 1,
    %ymode=log,
    xlabel={$E_b/N_0$ [\text{dB}]}, xlabel style={yshift=0.4em},
    ylabel={Complexity [\%]}, ylabel style={yshift=-0.75em},
    grid=both,
    ymajorgrids=true,
    xmajorgrids=true,
    grid style=dashed,
    mark options=solid,
    width=1\columnwidth, height=7cm,
    thick,
 %       xmin=3,
%        xmax=8,
%        ymin=1e-6,
    mark size=3,
    legend style={
      anchor={center},
      cells={anchor=west},
      mark options=solid,
      column sep= 2mm,
      font=\fontsize{7pt}{7.2}\selectfont,
    },
    legend to name=COMP_OPTIMAL_LLR,
    legend columns=2,
]

\addplot[
    color=CornflowerBlue,
    mark=square,
    thick,
    mark size=3,
]
table {
5.0 99.4402975
5.5 90.034
6.0 56.7265
6.5 30.4855
7.0 25.29925
};
\addlegendentry{$P_{high}$=5 $P_{low}$=3}
%4.5, 7

\addplot[
    color=blue,
    mark=o,
    thick,
    mark size=3,
]
table {
5.0 100
5.5 100
6.0 99.4702
6.5 82.5265
7.0 53.9415
};
\addlegendentry{$P_{high}$=5 $P_{low}$=4}
%7.5, 12

\addplot[
    color=BurntOrange,
	mark=triangle,
    thick,
    mark size=3,
]
table {
5.0     100	
5.5     100
6.0     100
6.5     100
7.0   	100
};
\addlegendentry{$P$=5}

\addplot[
    color=black,
    mark=x,
    thick,
    mark size=3,
]
table {
5       50
5.5     50
6       50
6.5     50
7       50
};
\addlegendentry{$P$=4}

\end{axis}
\end{tikzpicture}
   \end{minipage}
    \ref{BLER_OPTIMAL_LLR}
    \caption{BLER and complexity for IDA-Chase decoding with optimal threshold at $E_b/N_0$=6.5dB, $P_{high}$=5 $P_{low}$=\{3,4\}, and reference curves for Chase with $P$=\{4,5\}.}
    \label{fig:CHASELLR}
\end{figure}

IDA decoding, as introduced at the beginning of Section \ref{sec:IDA}, is applied to Chase decoding in Figure \ref{fig:CHASELLR}, for $P_{high}$=5 and $P_{low}$=\{3,4\}.
Thresholds are optimized for $E_b/N_0$=6.5dB, leading to $\gamma=$\{4.5,7.5\} and $\phi=$\{7,12\}. 
As expected, the results in \cite{IDASCL} transfer to other types of codes and decoding algorithms. 
IDA-Chase can guarantee significant complexity reduction at a tunable BLER cost: at $E_b/N_0$=6.5dB, $P_{low}=4$ reduces the complexity to $82.5\%$ and $P_{low}=3$ to $30.5\%$ with negligible BLER degradation with respect to $P=5$ and $P=4$, respectively. 
Moving away from the $E_b/N_0$ at which $\gamma$ and $\phi$ have been optimized, the BLER shifts to that of $P_{high}$ (lower $E_b/N_0$) or $P_{low}$ (higher $E_b/N_0$).

%\begin{figure}[t!]
%  \centering
%  \input{BLER_255_239_2.tikz}
%  \\
% \ref{BLER_reference}
%  \caption{CHASE curves with different nLRPs for BCH(255,239,2), ADC quantization (4,0).}
%  \label{fig:BLERref}
%\end{figure}

%\begin{figure}[t!]
%  \centering
%  \input{BLER_255_239_2_no_quant.tikz}
%  \\
% \ref{BLER_reference_noquant}
%  \caption{CHASE curves with different $P$ for BCH(255,239,2), no quantization.}
%  \label{fig:BLERref_noquant}
%\end{figure}
%
%\begin{figure}[t!]
%  \centering
%  \input{BLER_255_239_2_ORBGRAND.tikz}
%  \\
% \ref{BLER_ORBGRAND_noquant}
%  \caption{ORBGRAND curves with different nPat for BCH(255,239,2), no quantization.}
%  \label{fig:BLER_ORBGRAND_noquant}
%\end{figure}

%\begin{figure}[t!]
%  \centering
%  \input{BLER_255_239_2_OPTIMAL_DIFF_ORBGRAND.tikz}
%  \\
% \ref{BLER_OPTIMAL_ORB}
%  \caption{\fixme{BLER with LRP-based IDA ORBGRAND decoding with optimal threshold}{}}
%  \label{fig:BLERopt}
%\end{figure}

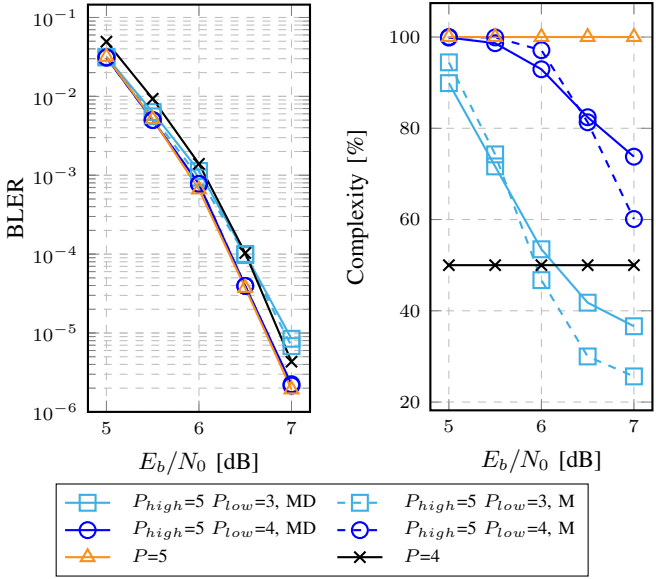
\begin{figure}[t!]
   \centering
    \begin{minipage}{.25\textwidth}
        \centering
%        		  \vspace{-12pt}
		  \begin{tikzpicture}
  \pgfplotsset{
    label style = {font=\fontsize{9pt}{7.2}\selectfont},
    tick label style = {font=\fontsize{7pt}{7.2}\selectfont}
  }

\begin{axis}[
	scale = 1,
    ymode=log,
    xlabel={$E_b/N_0$ [\text{dB}]}, xlabel style={yshift=0.4em},
    ylabel={BLER}, ylabel style={yshift=-0.75em},
    grid=both,
    ymajorgrids=true,
    xmajorgrids=true,
    grid style=dashed,
    mark options=solid,
    width=1\columnwidth, height=7cm,
    thick,
        ymin=1e-6,
    mark size=3,
    legend style={
      anchor={center},
      cells={anchor=west},
      mark options=solid,
      column sep= 2mm,
      font=\fontsize{7pt}{7.2}\selectfont,
    },
    legend to name=BLER_OPTIMAL_DIFF,
    legend columns=2,
]

\addplot[
    color=CornflowerBlue,
    mark=square,
    thick,
    mark size=3,
]
table {
5       0.0317662	 
5.5       0.0063139	 
6       0.0011317	 
6.5      0.00010058
7     8.4492e-006	 
};
\addlegendentry{$P_{high}$=5 $P_{low}$=3, MD}
%Diff>=2.2

\addplot[
    color=CornflowerBlue,
	dashed,
    mark=square,
    thick,
    mark size=3,
]
table {
5       0.0310945	
5.5       0.0055828	
6      0.00093433	
6.5     9.7272e-005	
7     6.8119e-006	
};
\addlegendentry{$P_{high}$=5 $P_{low}$=3, M}
%Abs>=3.0

\addplot[
    color=blue,
    mark=o,
    thick,
    mark size=3,
]
table {
5       0.0310945	 
5.5       0.0050015	 
6      0.00077572	 
6.5     3.9551e-005	 
7    2.2433e-006	
};
\addlegendentry{$P_{high}$=5 $P_{low}$=4, MD}
%Diff>3.6

\addplot[
    color=blue,
	dashed,
    mark=o,
    thick,
    mark size=3,
]
table {
5       0.0310945	
5.5       0.0050015	
6      0.00077572	
6.5     3.9551e-005	
7     2.1552e-006	
};
\addlegendentry{$P_{high}$=5 $P_{low}$=4, M}
%Abs>5

\addplot[
    color=BurntOrange,
	mark=triangle,
    thick,
    mark size=3,
]
table {
5.0     0.0316556	
5.5       0.0051414	
6.0    0.00066028	
6.5     3.7284e-005	
7.0   1.9278e-006	
};
\addlegendentry{$P$=5}

\addplot[
    color=black,
    mark=x,
    thick,
    mark size=3,
]
table {
5.0     0.0493583	
5.5       0.0093266	
6.0     0.0013959	
6.5      0.00010328	
7.0   4.3302e-006		 
};
\addlegendentry{$P$=4}

\end{axis}
\end{tikzpicture}
		        \end{minipage}%
    \begin{minipage}{0.25\textwidth}
        \centering
		  \begin{tikzpicture}
  \pgfplotsset{
    label style = {font=\fontsize{9pt}{7.2}\selectfont},
    tick label style = {font=\fontsize{7pt}{7.2}\selectfont}
  }

\begin{axis}[
	scale = 1,
    %ymode=log,
    xlabel={$E_b/N_0$ [\text{dB}]}, xlabel style={yshift=0.4em},
    ylabel={Complexity [\%]}, ylabel style={yshift=-0.75em},
    grid=both,
    ymajorgrids=true,
    xmajorgrids=true,
    grid style=dashed,
    mark options=solid,
    width=1\columnwidth, height=7cm,
    thick,
 %       xmin=3,
%        xmax=8,
%        ymin=1e-6,
    mark size=3,
    legend style={
      anchor={center},
      cells={anchor=west},
      mark options=solid,
      column sep= 2mm,
      font=\fontsize{7pt}{7.2}\selectfont,
    },
    legend to name=COMP_OPTIMAL_DIFF,
    legend columns=2,
]

\addplot[
    color=CornflowerBlue,
    mark=square,
    thick,
    mark size=3,
]
table {
5   89.89825
5.5 71.644
6   53.48425
6.5 41.782
7   36.61225
};
\addlegendentry{$P_{high}$=5 $P_{low}$=3, diff}
%Diff>=2.2

\addplot[
    color=CornflowerBlue,
    mark=square,
	dashed,
    thick,
    mark size=3,
]
table {
5   94.449625
5.5 74.2825
6   46.723
6.5 29.9965
7   25.621
};
\addlegendentry{$P_{high}$=5 $P_{low}$=3, abs}
%abs>=3.0

\addplot[
    color=blue,
    mark=o,
    thick,
    mark size=3,
]
table {
5   99.87562
5.5 98.6446
6   92.926
6.5 82.42
7    73.775
};
\addlegendentry{$P_{high}$=5 $P_{low}$=4, diff}
%Diff>=3.6

\addplot[
    color=blue,
	dashed,
    mark=o,
    thick,
    mark size=3,
]
table {
5 100
5.5 99.909975
6 97.1081
6.5 81.3125
7    60.1145
};
\addlegendentry{$P_{high}$=5 $P_{low}$=4, abs}
%Abs>5.0

\addplot[
    color=BurntOrange,
	mark=triangle,
    thick,
    mark size=3,
]
table {
5.0     100	
5.5     100
6.0     100
6.5     100
7.0   	100
};
\addlegendentry{$P$=5}

\addplot[
    color=black,
    mark=x,
    thick,
    mark size=3,
]
table {
5       50
5.5     50
6       50
6.5     50
7       50
};
\addlegendentry{$P$=4}

\end{axis}
\end{tikzpicture}
   \end{minipage}
    \ref{BLER_OPTIMAL_DIFF}
    \caption{BLER and complexity for M-IDA and MD-IDA-Chase decoding with optimal threshold at $E_b/N_0$=6.5dB, $P_{high}$=5 $P_{low}$=\{3,4\}, and reference curves for Chase with $P$=\{4,5\}.}
    \label{fig:CHASEOPT}
\end{figure}

%\begin{figure}[t!]
%  \centering
%  \input{Complexity_evo2.tikz}
%  \\
% \ref{Comp_OPTIMAL}
%  \caption{Complexity with LRP-based IDA CHASE decoding with optimal threshold, nLRP$_{high}$=6 nLRP$_{low}$=4.}
%  \label{fig:compOpt}
%\end{figure}

%\subsection{IDA-Chase decoding}

Figure \ref{fig:CHASEOPT} plots the BLER and the complexity percentage for a BCH(255,239,2) code decoded with Chase, M-IDA-Chase, and MD-IDA-Chase. 
IDA-Chase curves have been obtained with $P_{high}=5$, $P_{low}$=\{3,4\}, and the optimal $\gamma$ for $E_b/N_0$=6.5dB.
The selected thresholds minimize complexity while setting a maximum BLER equal to that of standard Chase with $P=4$ in the case of $P_{low}=3$, and of standard Chase with $P=5$ when $P_{low}=4$. 
It can be seen that a single optimized threshold ($\gamma_M$=5.0, $\gamma_{MD}$=3.6) for $P_{low}$=3 allows to match the BLER of Chase with $P=5$ for both M-IDA and MD-IDA.
At $E_b/N_0\ge$6.5dB, M-IDA yields lower complexity than MD-IDA, going as low as $60.1\%$, against the $73.8\%$ obtained by MD-IDA. 
With $P_{low}$=3, $\gamma_M$=3.0 and $\gamma_{MD}$=2.2 allow to match the BLER of standard Chase with $P=4$ at $E_b/N_0$=6.5dB. 
At lower $E_b/N_0$, as the percentage of time where $P_{high}$ is used increases, the complexity rises and BLER drifts towards that of Chase with $P=P_{high}=5$. 
An opposite effect is observed at higher $E_b/N_0$, where the BLER and complexity move toward that of Chase with $P=P_{low}=3$.
In terms of complexity M-IDA shows higher gains than MD-IDA throughout the majority of the $E_b/N_0$ range, with a minimum observed complexity of $25.6\%$ for M-IDA and $36.6\%$ for MD-IDA.
Compared to IDA-Chase, both M-IDA and MD-IDA yield comparable complexity reduction, even performing better in case of M-IDA with $P_{low}=3$. 

\begin{figure}[t!]
   \centering
    \begin{minipage}{.25\textwidth}
        \centering
  \begin{tikzpicture}[spy using outlines={circle, magnification=4, size=1.7cm, connect spies, transform shape}]
  \pgfplotsset{
    label style = {font=\fontsize{9pt}{7.2}\selectfont},
    tick label style = {font=\fontsize{7pt}{7.2}\selectfont}
  }

\begin{axis}[
	scale = 1,
    ymode=log,
    xlabel={$E_b/N_0$ [\text{dB}]}, xlabel style={yshift=0.4em},
    ylabel={BLER}, ylabel style={yshift=-0.75em},
    grid=both,
    ymajorgrids=true,
    xmajorgrids=true,
    grid style=dashed,
    mark options=solid,
    width=1\columnwidth, height=7cm,
    thick,
 %       xmin=3,
%        xmax=8,
 %       ymin=1e-6,
    mark size=3,
    legend style={
      anchor={center},
      cells={anchor=west},
      mark options=solid,
      column sep= 2mm,
      font=\fontsize{7pt}{7.2}\selectfont,
    },
    legend to name=BLER_OPTIMAL_ORB,
    legend columns=2,
]

\addplot[
    color=blue,
    mark=o,
    thick,
    mark size=3,
]
table {
5           0.154	
5.5         0.04771	
6        0.012953	
6.5       0.0026767	
7      0.0003317	
};
\addlegendentry{nPat$_{high}$=500, nPat$_{low}$=252, MD}
%Diff>=8.2, weight_low=18

\addplot[
    color=BurntOrange,
    mark=triangle,
    thick,
    mark size=3,
]
table {
5           0.154	
5.5         0.04771	
6         0.01245	
6.5       0.0023929	
7      0.00031722			 
};
\addlegendentry{nPat=500}

\addplot[
    color=blue,
	dashed,
    mark=o,
    thick,
    mark size=3,
]
table {
5           0.154	
5.5         0.04771	
6         0.01245	
6.5       0.0023929	
7      0.00035037	 
};
\addlegendentry{nPat$_{high}$=500, nPat$_{low}$=252, M}
%Abs>=10, weight_low=18

\addplot[
    color=black,
	mark=x,
    thick,
    mark size=3,
]
table {
5           0.162	
5.5        0.049164	
6         0.01339	
6.5       0.0024629	
7      0.00038264	
};
\addlegendentry{nPat=446}

\addplot[
    color=CornflowerBlue,
    mark=square,
    thick,
    mark size=3,
]
table {
5           0.154	
5.5         0.04771		
6        0.013503	
6.5       0.0028609		
7      0.00040687	
};
\addlegendentry{nPat$_{high}$=500, nPat$_{low}$=168, MD}
%Diff>=7.6, weight_low=16

\addplot[
    color=gray,
	mark=+,
    thick,
    mark size=3,
]
table {
5             0.2	
5.5           0.067	
6        0.018896	
6.5       0.0032973	
7      0.00063684	
};
\addlegendentry{nPat=252}

\addplot[
    color=CornflowerBlue,
	dashed,
    mark=square,
    thick,
    mark size=3,
]
table {
5           0.154	  
5.5         0.04771	  	
6         0.01245	  
6.5       0.0023929	  	
7      0.00039113	  
};
\addlegendentry{nPat$_{high}$=500, nPat$_{low}$=168, M}
%Diff>=9.2, weight_low=16

\addplot[
    color=Bittersweet,
	dashed,
    thick,
    mark size=3,
]
table {
5           0.223	
5.5           0.084	
6        0.021413	
6.5       0.0050368	
7      0.00091483	
};
\addlegendentry{nPat=168}

%\addplot[
%    color=magenta,
%	%mark=magenta,
%    thick,
%    mark size=3,
%]
%table {
%5           0.184	
%5.5           0.056	
%6        0.015408	
%6.5       0.0029307	
%7        0.000561	
%};
%\addlegendentry{nPat=306}
%
%\addplot[
%    color=gray,
%	%mark=triangle,
%    thick,
%    mark size=3,
%]
%table {
%5           0.172	
%5.5           0.053	
%6        0.013466	
%6.5       0.0026919	
%7      0.00039549	
%};
%\addlegendentry{nPat=369}

\spy [black] on (2.71,0.55) in node [left] at (1.85,1.5);

\end{axis}
\end{tikzpicture}
     \end{minipage}%
    \begin{minipage}{0.25\textwidth}
        \centering
		  \begin{tikzpicture}
  \pgfplotsset{
    label style = {font=\fontsize{9pt}{7.2}\selectfont},
    tick label style = {font=\fontsize{7pt}{7.2}\selectfont}
  }

\begin{axis}[
	scale = 1,
    %ymode=log,
    xlabel={$E_b/N_0$ [\text{dB}]}, xlabel style={yshift=0.4em},
    ylabel={Complexity [\%]}, ylabel style={yshift=-0.75em},
    grid=both,
    ymajorgrids=true,
    xmajorgrids=true,
    grid style=dashed,
    mark options=solid,
    width=1\columnwidth, height=7cm,
    thick,
 %       xmin=3,
%        xmax=8,
%        ymin=1e-6,
    mark size=3,
    legend style={
      anchor={center},
      cells={anchor=west},
      mark options=solid,
      column sep= 2mm,
      font=\fontsize{7pt}{7.2}\selectfont,
    },
    legend to name=COMP_OPTIMAL_ORB,
    legend columns=2,
]

\addplot[
    color=blue,
    mark=o,
    thick,
    mark size=3,
]
table {
5 100
5.5 99.43208
6 95.2070528
6.5 87.480464
7 82.543776
};
\addlegendentry{nPat$_{high}$=500, nPat$_{low}$=252, diff}
%Diff>=8.2, weight_low=18

\addplot[
    color=blue,
	dashed,
    mark=o,
    thick,
    mark size=3,
]
table {
5 100
5 100
6 100
6 99.86944288
7 89.19712
};
\addlegendentry{nPat$_{high}$=500, nPat$_{low}$=252, abs}
%abs>=10, weight_low=18

\addplot[
    color=CornflowerBlue,
    mark=square,
    thick,
    mark size=3,
]
table {
5 99.9336
5.5 98.416028
6 92.020712
6.5 81.626456
7 75.897464
};
\addlegendentry{nPat$_{high}$=500, nPat$_{low}$=168}
%Diff>=7.6, weight_low=18

\addplot[
    color=CornflowerBlue,
	dashed,
    mark=square,
    thick,
    mark size=3,
]
table {
5   100
5.5 100
6   100
6.5 98.7002864
7   77.951216
};
\addlegendentry{nPat$_{high}$=500, nPat$_{low}$=168}
%abs>=, weight_low=16

\addplot[
    color=BurntOrange,
    mark=triangle,
    thick,
    mark size=3,
]
table {
5      100
5.5    100
6      100
6.5    100	
7      100	 
};
\addlegendentry{nPat=500}

\addplot[
    color=black,
	mark=x,
    thick,
    mark size=3,
]
table {
5     89.2 
5.5   89.2 
6     89.2 
6.5   89.2 
7     89.2 
};
\addlegendentry{nPat=446}

%\addplot[
%    color=BurntOrange,
%	%mark=triangle,
%    thick,
%    mark size=3,
%]
%table {
%5      50.4
%5.5    50.4
%6      50.4
%6.5    50.4
%7      50.4
%};
%\addlegendentry{nPat=252}

%\addplot[
%    color=magenta,
%	%mark=magenta,
%    thick,
%    mark size=3,
%]
%table {
%5           0.184	
%5.5           0.056	
%6        0.015408	
%6.5       0.0029307	
%7        0.000561	
%};
%\addlegendentry{nPat=306}
%
%\addplot[
%    color=gray,
%	%mark=triangle,
%    thick,
%    mark size=3,
%]
%table {
%5           0.172	
%5.5           0.053	
%6        0.013466	
%6.5       0.0026919	
%7      0.00039549	
%};
%\addlegendentry{nPat=369}

\end{axis}
\end{tikzpicture}
   \end{minipage}
    \ref{BLER_OPTIMAL_ORB}
    \caption{BLER and complexity for M-IDA and MD-IDA-ORBGRAND decoding with optimal threshold at $E_b/N_0$=7dB, nPat$_{high}$=500 nPat$_{low}$=252, and reference curves for ORBGRAND with nPat=\{500,446,252,168\}.}
    \label{fig:ORBOPT}
\end{figure}
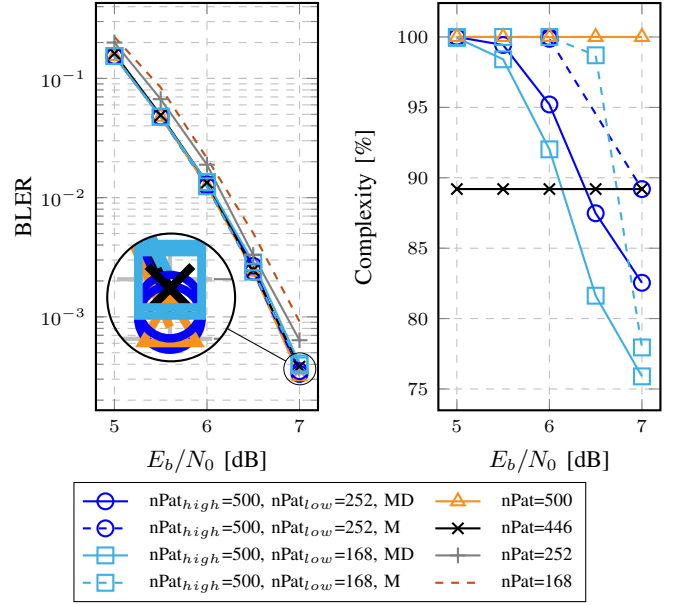

%\subsection{IDA-ORBGRAND decoding}
Figure \ref{fig:ORBOPT} portrays the BLER of the same code decoded with ORBGRAND, M-IDA-ORBGRAND, and MD-IDA-ORBGRAND. 
IDA-ORBGRAND curves consider nPat$_{high}$=$500$ ($w_{{\rm nPat}_{high}}$=22) and nPat$_{low}$=\{252,168\} ($w_{{\rm nPat}_{high}}$=\{18,16\}), and have been obtained with $\gamma$ optimized for $E_b/N_0$=7dB.
With nPat$_{low}$=252 ($\gamma_M$=10, $\gamma_{MD}$=8.2), the BLER of IDA-ORBGGRAND matches that of ORBGRAND with nPat=500, with a complexity of $82.5\%$ for MD-IDA and of $89.2$ for M-IDA, while nPat$_{low}$=168 ($\gamma_{M}$=9.2, $\gamma_{MD}$=7.6) allows to match the BLER of ORBGRAND with nPat=446, further lowering the complexity to $75.9\%$ for MD-IDA and to $78\%$ for M-IDA.
The reference curves for ORBGRAND with nPat=\{252,168\} show the potential BLER degradation of IDA-ORBGRAND: however, the selection of nPat$_{low}$ is able to minimize the BLER degradation while granting substantial complexity reduction.

It can be observed that complexity reduction is dominated by M-IDA for IDA-Chase, while MD-IDA performs better in case of IDA-ORBGRAND.
This is due to the fact that in M-IDA-Chase, the observation of $|\tilde{y}_{P_{high}-1}|$ is very informative, since switching from $P_{high}$ to $P_{low}$ assumes that $|\tilde{y}_{P_{high}-1}|$ and at most $|\tilde{y}_{P_{high}-2}|$ are removed from the least reliable LLRs.
On the other hand, due to the bit flipping patterns of ORBGRAND being based on logistic weights, switching from nPat$_{high}$ to nPat$_{low}$ excludes patterns that involve a multitude of LLRs, of which not much can be inferred by observing $|\tilde{y}_{w_{{\rm nPat}_{high}}-1}|$ only. 
Instead, MD-IDA-ORBGRAND gives a sense of the clustering of the least reliable LLRs, resulting in a more reliable decision process.

%First of all, the bit flipping patterns created by Chase consider all combinations of the $P$ least reliable LLRs: reducing $P$ by removing the value of $|\tilde{y}_i|$ with $i>P_{low}$ implies that all bits $i$ are assumed correct, but no assumption is made on the remaining $|\tilde{y}_{P_{low}}|$ positions.
%Moreover, in the presented results for IDA-Chase $P_{low}$ is chosen by observing $\tilde{y}_{P_{high}}$, and $P_{low}=P_{high}-\{1,2\}$. 
%As $\tilde{y}_{P_{high}}$ is directly observed, its removal from the $P$ least reliable positions is an informed decision that causes very little BLER degradation, but any other removed $\tilde{y}_i$ is not directly observed.
%As the difference between $P_{high}$ and $P_{low}$ increases, the decision to use $P_{low}$ is based on less information and is thus less precise.

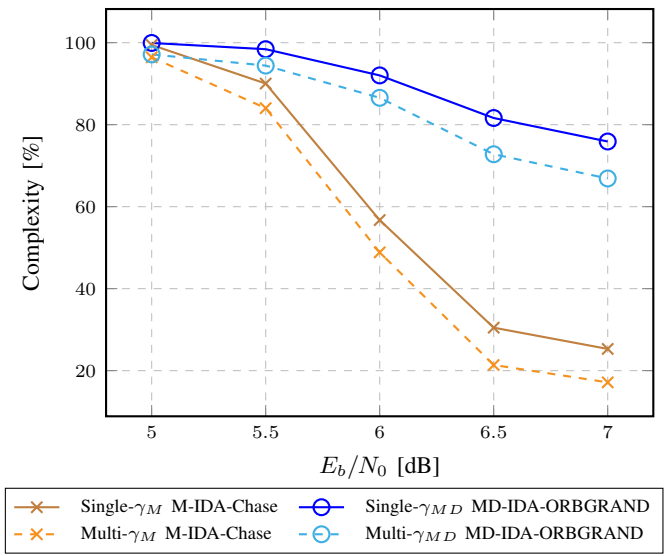
\begin{figure}[t!]
  \centering
  \begin{tikzpicture}
  \pgfplotsset{
    label style = {font=\fontsize{9pt}{7.2}\selectfont},
    tick label style = {font=\fontsize{7pt}{7.2}\selectfont}
  }

\begin{axis}[
	scale = 1,
    %ymode=log,
    xlabel={$E_b/N_0$ [\text{dB}]}, xlabel style={yshift=0.4em},
    ylabel={Complexity [\%]}, ylabel style={yshift=-0.75em},
    grid=both,
    ymajorgrids=true,
    xmajorgrids=true,
    grid style=dashed,
    mark options=solid,
    width=1\columnwidth, height=7cm,
    thick,
 %       xmin=3,
%        xmax=8,
%        ymin=1e-6,
    mark size=3,
    legend style={
      anchor={center},
      cells={anchor=west},
      mark options=solid,
      column sep= 2mm,
      font=\fontsize{7pt}{7.2}\selectfont,
    },
    legend to name=COMP_MULTI,
    legend columns=2,
]

\addplot[
    color=brown,
    mark=x,
    thick,
    mark size=3,
]
table {
5.0 99.4402975
5.5 90.034
6.0 56.7265
6.5 30.4855
7.0 25.29925
};
\addlegendentry{Single-$\gamma_M$ M-IDA-Chase}

\addplot[
    color=blue,
    mark=o,
    thick,
    mark size=3,
]
table {
5 99.9336
5.5 98.416028
6 92.020712
6.5 81.626456
7 75.897464
};
\addlegendentry{Single-$\gamma_{MD}$ MD-IDA-ORBGRAND}
%Diff>=8.2, weight_low=18

\addplot[
    color=BurntOrange,
    mark=x,
    dashed,
    thick,
    mark size=3,
]
table {
5.0 96.4402975
5.5 84.034
6.0 48.85206
6.5 21.3761
7.0 17.14337
};
\addlegendentry{Multi-$\gamma_M$ M-IDA-Chase}

\addplot[
    color=CornflowerBlue,
    mark=o,
    dashed,
    thick,
    mark size=3,
]
table {
5 97.141
5.5 94.416028
6 86.5561
6.5 72.7878
7 66.88052
};
\addlegendentry{Multi-$\gamma_{MD}$ MD-IDA-ORBGRAND}
%Diff>=8.2, weight_low=18

%\addplot[
%    color=BurntOrange,
%	%mark=triangle,
%    thick,
%    mark size=3,
%]
%table {
%5      50.4
%5.5    50.4
%6      50.4
%6.5    50.4
%7      50.4
%};
%\addlegendentry{nPat=252}

%\addplot[
%    color=magenta,
%	%mark=magenta,
%    thick,
%    mark size=3,
%]
%table {
%5           0.184	
%5.5           0.056	
%6        0.015408	
%6.5       0.0029307	
%7        0.000561	
%};
%\addlegendentry{nPat=306}
%
%\addplot[
%    color=gray,
%	%mark=triangle,
%    thick,
%    mark size=3,
%]
%table {
%5           0.172	
%5.5           0.053	
%6        0.013466	
%6.5       0.0026919	
%7      0.00039549	
%};
%\addlegendentry{nPat=369}

\end{axis}
\end{tikzpicture}
  \\
 \ref{COMP_MULTI}
  \caption{Complexity for single-threshold and multi-threshold M-IDA-Chase and MD-IDA-ORBGRAND decoding with optimal threshold at $E_b/N_0$=6.5 and 7dB, resepectively.}
  \label{fig:multi}
\end{figure}

As with IDA-SCL in \cite{IDASCL}, it is possible to use multiple concurrent thresholds $\gamma$ to decide on different possible $P_{low}$ (nPat$_{low}$), in order to further decrease the complexity. 
In this case, complexity is computed as
\begin{equation}\label{eq:comp_ML_Chase}
100 \cdot \sum_{i=0}^{P_{high}} \delta_i \cdot \frac{2^i}{2^{P_{high}}}
\end{equation}
for Chase, and as
\begin{equation}\label{eq:comp_ML_ORB}
100 \cdot \sum_{i=0}^{w_{{\rm nPat}_{high}}} \delta_i \cdot \frac{{\rm nPat}_i}{{\rm nPat}_{high}},
\end{equation}
in case of ORBGRAND, where $\delta_i$ is the fraction of times that $P_{low}=2^i$ (nPat$_{low}$=nPat$_i$) is selected. 
Figure \ref{fig:multi} shows the complexity for M-IDA-Chase and MD-IDA-ORBGRAND in case of $P_{low}$=\{1,2,3,4\} and nPat$_{low}$=\{369,306,252,168\}. 
Thresholds are selected to maximize complexity reduction at $E_b/N_0$=6.5dB and 7dB, respectively, while the BLER matches that of Chase with $P$=4 and ORBGRAND with nPat=446. 
It can be seen that the use of multiple thresholds allows to further reduce the decoding complexity, reaching as low as $17\%$ in case of M-IDA-Chase. 
%As each threshold couple misidentifies a potentially different set of cases, the BLER degradation caused by each $L_{low}$ is partially accumulated, leading to unacceptable performance. 
%For this reason, the optimal thresholds identified for each $L_{low}$ need to be decreased. 

IDA decoding can be implemented in practical decoders at the cost of some threshold comparisons and counters, whose number depends on how many LLRs are received from the channel at each clock cycle. 
M-IDA and MD-IDA decoding require the identification of some of the minimum elements of $|\mathbf{y}|$: this operation can be quite costly, depending on the number of elements in $\mathbf{y}$ and the desired minima. 
However, for algorithms like Chase and ORBGRAND, this information is part of the standard decoding process, and thus comes at no extra cost. 
In these cases, M-IDA can be implemented at the cost of a single comparator, while MD-IDA requires a subtractor and a comparator; the latency of these computations is independent of the code length.

\section{Conclusion}\label{sec:conc}
In this work, two low-complexity incarnations of input-distribution-aware (IDA)  decoding \cite{IDASCL} were proposed, called M-IDA and MD-IDA. 
They exploit the partial sorting of channel LLRs that is inherent to some decoding algorithms, like Chase and ORBGRAND, to dynamically reduce the number of parallel decoding attempts necessary for successful decoding, leading to energy saving at a negligible implementation cost. 
Moreover, the original IDA decoding approach was extended to different types of codes and decoding algorithms. 
%The IDA approach allows to deactivate part of a decoder before decoding, thus leading to energy saving at a negligible implementation cost. 
The proposed techniques are shown to reduce the decoding complexity of Chase and ORBGRAND down to $17\%$ and $67\%$, respectively, while meeting a performance target.

%\bibliographystyle{IEEEbib}
%\bibliography{IEEEabrv,refs}

\end{document}